\documentstyle[twoside,fleqn,espcrc2,epsf]{article}
\def\dsum{\displaystyle\sum}
\def\Re{{\rm Re}\,}
\def\Tr{{\rm Tr}\,}

\title{Effect of Improving the Lattice Gauge Action on QCD Topology
\thanks{Presented by J.~Grandy.}}

\author{J.~Grandy\address{L-170, Lawrence Livermore National
Laboratory, P.~O.~Box 808, Livermore, CA 94551}
and G.~Kilcup,\address{Physics Department, 174 West 18th Avenue, 
Ohio State University, Columbus, OH 43210}}

\begin{document}

\begin{abstract}
We use lattice topology as a laboratory to compare the Wilson action
(WA) with the Symanzik-Weisz (SW) action constructed from a
combination of $(1\times1)$ and $(1\times2)$ Wilson loops, and the
estimate of the renormalization trajectory (RT)\cite{gupta87} from a
renormalization group transformation (RGT) which also includes higher
representations of the $(1\times1)$ loop.  Topological charges are
computed using the geometric (L\"uscher's) and plaquette methods on
the uncooled lattice, and also by using cooling to remove ultraviolet
artifacts. We show that as the action improves by approaching the RT,
the topological charges for individual configurations computed using
these three methods become more highly correlated, suggesting that
artificial lattice renormalizations to the topological susceptibility
can be suppressed by improving the action.

\end{abstract}

\maketitle

\section{Improved Actions}
In lattice QCD, one of the fundamental challenges is to minimize the
errors caused by discretizing space-time.  This is accomplished
through a combination of advances in computer technology, and advances
in the formulation of methods to solve the problem computationally,
including the development of improved numerical algorithms.  We show
an example of how an improved gauge field action can be used to
suppress artificial lattice contributions to physical measurements.

We consider gluon actions that are constructed in a gauge invariant
fashion, from a combination of Casimir invariants of closed Wilson
loops.  In principle, a lattice action of this type can consist of an
arbitrary sum of Wilson loops, but a truncation to a small set of
localized loops is necessary due to computational expense.  We
study actions constructed from $(1\times1)$ and
$(1\times2)$ loops: 

\setlength{\tabcolsep}{0.00pc}
\vspace{0.25cm}
\begin{tabular}{@{\hspace{-0.5pc}}lcl@{\hspace{0.0pc}}}
\multicolumn{2}{@{\hspace{-0.5pc}}l@{\hspace{0.0pc}}}{$S=
         K^{1\times1}   $}&$ \dsum  \Re\Tr W^{(1 \times 1)}  $ \cr
$ + $&$ K^{1\times2}            $&$   \dsum \Re\Tr W^{(1 \times 2)} $ \cr
$ + $&$ K_6^{1 \times 1} $&$   \dsum \Re[ 
{3 \over 2} (\Tr W^{(1 \times 1)} )^2
- {1 \over 2} \Tr W^{(1 \times 1)} ] $ \cr
$ + $&$ K_8^{1 \times 1} $&$   \dsum [ {9 \over 8} \Tr W^{(1 \times 1)} |^2
                         - {1 \over 8} ]$  \cr
\end{tabular}
\vspace{0.25cm}
where the actions and coefficients, in order of increasing improvement
(approximation to the renormalization trajectory) are given by:

\vspace{0.5cm}
\setlength{\tabcolsep}{0.60pc}
\begin{tabular}{@{\hspace{-0.5pc}}lcccc@{\hspace{0.0pc}}}
Action &$ K^{1\times1} $&$ K^{1\times2}  $&$K_6^{1 \times 1}$&$ 
K_8^{1 \times 1}         $\cr
WA  &$  1 $&$  0  $&$  0  $&$  0  $ \cr
SW  &$ 5/3$&$-1/12$&$  0  $&$  0  $ \cr
RGT  &$  k $&$  -0.04k   $&$  -0.12k   $&$
  -0.12k  $ .  \cr
\end{tabular}
\vspace{0.5cm}

To compare these three actions, we generate four ensembles:

\vspace{0.5cm}
\setlength{\tabcolsep}{0.45pc}
\begin{tabular}{@{\hspace{-0.5pc}}lrrrr@{\hspace{0.0pc}}}
Action & \multicolumn{1}{c}{Size} &$\beta$&
\multicolumn{1}{r}{$\approx\beta_{Wil}$}&$\ \ N$  \cr
Wilson &$ (16^3\times40) $ &$\ \  6.0 $&$ 6.0\,\,\, $&$ 35 $\cr
SW     &$ (16^3\times32) $ &$\ \  4.2 $&$ 5.8\,\,\, $&$ 36 $\cr
SW     &$ (16^3\times32) $ &$\ \  4.43$&$ 6.0\,\,\, $&$ 40 $\cr
RGT    &$ (18^3\times36) $ &$k=10.58$&$ 6.0\,\,\, $&$ 28 $\cr
\end{tabular}
\vspace{0.5cm}

The two SW ensembles allow us to study the effect of increasing
$\beta$, we have used estimates of corresponding Wilson action $\beta$
from the deconfining phase transition temperature calculation by Cella
{\it et al.\/}\cite{Cella}.  Since we have used a modest number of
configurations in each case, we focus on the qualitative comparison
between Wilson and improved actions.  Further calculations with a
larger number of lattices would be needed for quantitative studies,
for example, to determine the consistency and scaling of
$\chi_t/m_\rho$.

\section{Topology: Comparing Actions}

Lattice topology provides a test case for comparing various gauge
field actions.  There are several prescriptions for measuring
topological charge
$$ Q = {1\over32\pi^2} \int d^4x F(x) \tilde{F}(x) $$ on the lattice, and
each prescription is subject to a different set of lattice cutoffs and
renormalizations which affect the measurement of the topological
susceptibility $\chi_t=\left\langle Q^2\right\rangle/V$.  In the
plaquette method the topological density $F(x)\tilde{F}(x)$ is
constructed from a product of lattice $(1\times1)$ Wilson loops.  This
method in general gives noninteger values of the topological charge,
and is affected by large multiplicative and additive lattice
renormalizations\cite{mixing}.  The geometric method\cite{Luscher}
does guarantee an integer topological charge (except for
``exceptional'' configurations) but is not guaranteed to obey physical
scaling in the continuum limit, and is in fact known to violate
scaling for the Wilson action\cite{Gockeler89}.  Low-action
dislocations which can be suppressed by improving the
action\cite{Gockeler89} contaminate the geometric $\chi_t$.  

In the cooling prescription, ultraviolet fluctuations in the fields
are removed by locally minimizing the action in successive sweeps,
isolating instanton-like configurations.  After cooling, a single
instanton configuration spanning several lattice spacings has a
computed charge of nearly one using either the geometric or plaquette
formula; we therefore apply the plaquette formula to the cooled
configurations to obtain a value for $Q$.  Lattice artifacts are very
different among these methods, and we can in general get different
results for plaquette ($Q_p$), the geometric ($Q_g$), and the cooling
($Q_c$) topological charges computed on the same original
configuration.  For improved actions, we expect lattice artifacts such
as dislocations to be suppressed, therefore we test this prediction by
comparing the different topological charge methods {\it with each
other\/}.

The cooling prescription actually encompasses a family of cooling
algorithms.  Typically one cools by selecting a link $U$ to minimize
some action $S_c$, and since cooling is merely used as a tool to
isolate instantons, there is no reason to tie $S_c$ to the Monte Carlo
gauge action $S$.  The cooling algorithms $S_c$ we consider here are
linear combinations of Wilson loops with coefficients $c_{(1\times1)}$
and $c_{(1\times2)}$, and since action is minimized only the ratio
$r_{12} = c_{(1\times2)}/c_{(1\times1)}$ is significant.  The cooling
algorithm with $r_{12}=-0.05$ removes the leading scaling violation
from the classical instanton action, and we also include cooling
algorithms with $r_{12}=0$ and $r_{12}=-0.093$, which has been derived
from a linear weak coupling approximation to the RGT action, for
comparison.  For the case $r_{12}=0$, the lack of barrier to a
decrease in the instanton size causes the instanton to disappear by
implosion during the cooling process, and for $r_{12}=-0.093$ a large
instanton expands until halted by the boundary.  We cool for $200$
sweeps for all three algorithms, and the comparison between these
three in an indication of the systematic effect of picking some
particular means of cooling.  We note that with $200$ sweeps of Wilson
cooling most of the topological charges are retained, since the large
instantons haven't had enough time to implode.  In general, we do not
see any effect from the selection of the cooling algorithm, except
perhaps in one ensemble.

\begin{figure}[t]
\epsfxsize=7.0cm \epsfbox{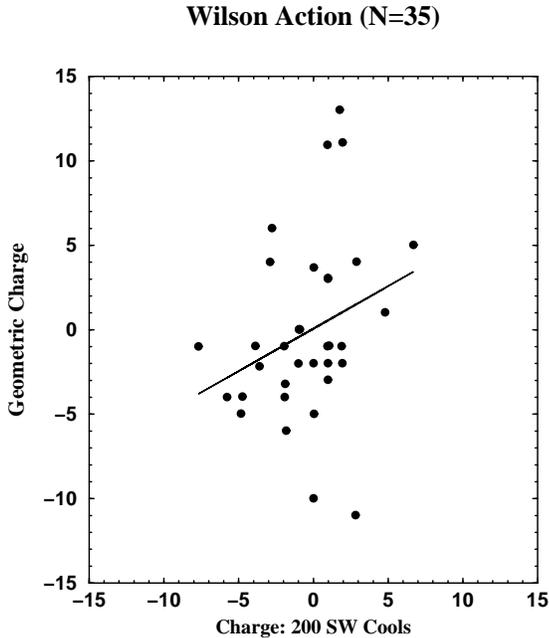}
\caption{Comparison of geometric and cooled topological charges for
Wilson action lattices at $a \approx 0.1{\rm fm}$.  Each point
represents one configuration, with the cooled charge as the absicca
and the geometric charge on the uncooled lattice as the ordinate.  The
least squares linear fit is shown.  Due to close overlaps there appear
to be fewer that $35$ points.}
\label{fig:qgcwa}
\end{figure}

\begin{figure}[t]
\epsfxsize=7.0cm \epsfbox{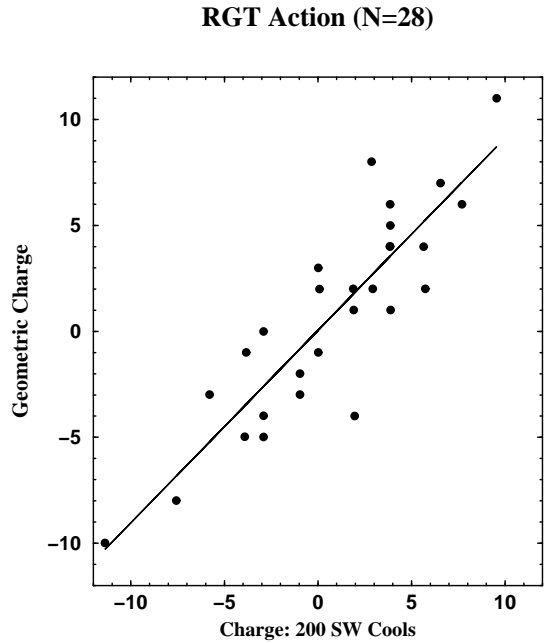}
\caption{Comparison of geometric and cooled topological charges for RGT
action lattices at $a \approx 0.1{\rm fm}$.}
\label{fig:qgcra}
\end{figure}

\section{Results}

As described above, we compute $Q_p$, $Q_g$, and $Q_c$ on all of our
lattices.  We show two scatter plots (Figures \ref{fig:qgcwa},
\ref{fig:qgcra}) highlighting the discrepancy between $Q_g$ and $Q_c$.
The best fit line is constructed through the points on a scatter plot.
The slope of this line is an estimate of the ratio of multiplicative
renormalizations, and should be close to $1$ since both the geometric
and cooling methods give integer charges.  The correlation
$$z_{gc} = \frac{\left\langle\left( Q_g - \bar{Q}_g\right)
\left( Q_c - \bar{Q}_c\right)\right\rangle }
{\sqrt{\left\langle\left(Q_g-\bar{Q}_g\right)^2\right\rangle
\left\langle\left(Q_c-\bar{Q}_c\right)^2\right\rangle}  } $$
between $Q_g$ and $Q_c$ is a measure of random additive artifacts seen
by one method but not the other, such as dislocations which disappear
during cooling therefore contributing only to $Q_g$ but not to $Q_c$.
The scatter plots show a strong correlation between $Q_g$ and $Q_c$
for the RGT action, and a far weaker correlation for the WA,
suggesting that the effect of lattice artifacts on topological charge
is far less pronounced for the improved action.

\begin{figure}[t]
\epsfxsize=7.2cm \epsfbox{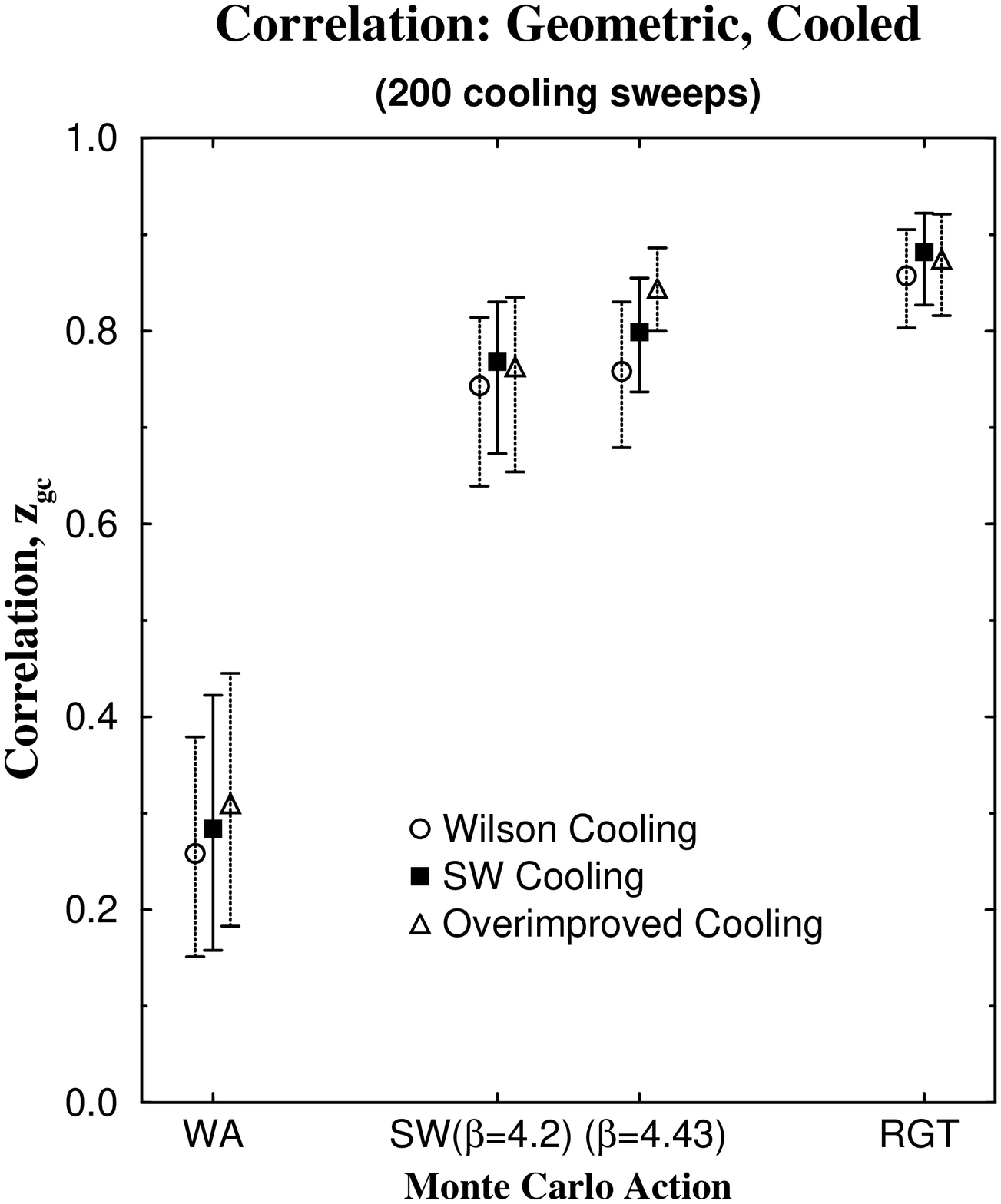}
\vspace{-0.5cm}
\caption{Correlation $z_{gc}$ between geometric and cooled topological
charges, for WA, SW, and RGT ensembles.  Results for three different
cooling methods are shown.}
\label{fig:zgc}
\end{figure}

\begin{figure}[t]
\epsfxsize=7.2cm \epsfbox{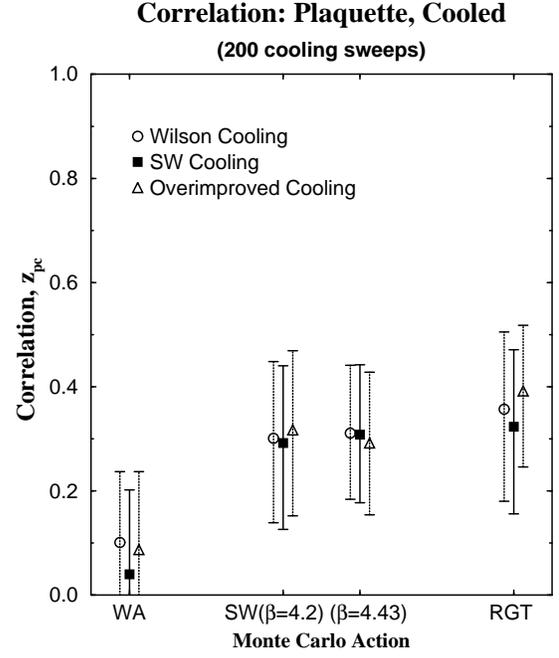}
\vspace{-0.5cm}
\caption{Statistical correlation $z_{pc}$ between plaquette and cooled
topological charges, for same ensembles, three cooling methods.}
\label{fig:zpc}
\end{figure}

\begin{table}[b]
\caption{Correlations, SW Cooling Linear Fits}
\setlength{\tabcolsep}{0.39pc}
\begin{tabular}{@{\hspace{0.2pc}}lllll@{\hspace{0.0pc}}}
Corr. &  \multicolumn{1}{c}{WA} & \multicolumn{1}{c}{SW, $4.2$} & 
\multicolumn{1}{c}{SW, $4.43$} & \multicolumn{1}{c}{RGT}  \cr
$z_{gc}  $&$  0.28(13) $&$  0.77(8) $&$   0.80(6) $&$   0.88(5) $\cr
$z_{pc}  $&$  0.04(16) $&$  0.29(16) $&$  0.31(13) $&$  0.32(16)$\cr
\end{tabular}
\label{tab:corr}
\end{table}
\vspace{0.25cm}

In Fig.~\ref{fig:zgc} we show a comparison between the WA, SW, and RGT
actions of the correlation between $Q_g$ and $Q_c$.
Fig.~\ref{fig:zpc} similarly shows the correlation between $Q_p$ and
$Q_c$, computed in the same manner as $z_{gc}$, and numerical values
are in Table \ref{tab:corr}.  The SW action serves as an intermediate
point between the other two actions, since the RGT action represents a
better estimate of the renormalization trajectory than the WA and SW
actions.  It is unclear whether the spread in $z_{gc}$ at $\beta=4.43$
is due to any systematic effect of the cooling algorithm.  It is
possible that for better improved actions, where the exponential
suppression of dislocations is greater than for the S-W action,
increasing beta will have a more profound effect than we have seen
here for the S-W action.  For the plaquette method, we have shown
previously\cite{lat94} that the multiplicative $Z_P$ becomes less
severe as the action improves, and the increased correlation $z_{pc}$
suggests that the additive renormalization also decreases. In
addition, improving the action is far more effective than increasing
$\beta$ for suppressing lattice artifacts.

Future calculations can include a more comprehensive study of other
improved actions.  Other methods for $\chi_t$, including the fermionic
method\cite{SmVink}, and an indirect measurement by calculating the
$\eta'$ mass\cite{Tsuk94,lat95}, should also be tested.  Having
established a correlation between cooled and uncooled topology and
located individual instantons, we are now prepared to investigate
Shuryak's picture of a dilute instanton gas, and the influence of
instantons on hadronic physics by working directly on uncooled
lattices. 

\noindent{\bf Acknowledgement.} The calculations were
performed at the Ohio (OSC) and National Energy Research
(NERSC) supercomputer centers, and the Advanced Computing Laboratory.


\begin{thebibliography}{\hspace{6pt}}
\bibitem{gupta87} Gupta, Patel, {\it Phys.Lett.} {\bf B183}, 193 (1987).
\bibitem{Cella} Cella, {\it Phys. Lett.} {\bf  B333}, 457 (1994).
\bibitem{mixing} Christou, {\it Phys.Rev.} {\bf D53}, 2619 (1996).
\bibitem{Luscher} L\"{u}scher, {\it Comm.Math.Phys.}{\bf 85}, 39 (1982).
\bibitem{Gockeler89} G\"{o}ckeler, {\it Phys.Lett.}{\bf B233}, 192 (1989).
\bibitem{lat94} Grandy, J.~and R.~Gupta, {\it Nucl.~Phys.~B
(Proc. Suppl.)} {\bf 42}, 246 (1995).
\bibitem{SmVink} Smit \& Vink, {\it Phys.Lett.} {\bf B194}, 433 (1987).
\bibitem{Tsuk94} Kuramashi, {\it Phys.Rev.Lett.\/} {\bf 72}, 3448 (1994).
\bibitem{lat95} Kilcup, {\it Nucl.~Phys.~B (Proc. Suppl.)} {\bf 47},
358 (1996).
\end{thebibliography}
\end{document}